# Propagating rotational jump events drive liquid-liquid transition in supercooled liquid water

**Biman Jana, Rakesh Saran Singh and Biman Bagchi***

Solid State and Structural Chemistry Unit,

Indian Institute of Science, Bangalore-12, India

*Email: bbagchi@sscu.iisc.ernet.in

*Abstract*

We observe, at low temperature, the appearance of propagating events that originate from the inter-conversion between adjacent four and five coordinated water molecules in supercooled liquid water, resulting in the migration of the coordination number *five* in a string-like fashion, creating rotational jumps along the way. The length of the connected events increases with lowering temperature. Each propagation event is terminated by a 3-coordinated species, present in a small number at large supercooling, which interacts cooperatively with 5-coordinated molecules to annihilate both the species. We find that these growing dynamical correlations manifest in a divergent-like growth of a non-linear density response function, $\chi_4(t)$, which is given by a four point time correlation function (FPTCF). The locus of the maximum of $\chi_4(t)$, when plotted against the time $t^*$ of maximum, exhibits a *sharp peak precisely at the temperature* where the static response functions (specific heat, isothermal compressibility) also show similar sharp, divergent-like, peak. While the decay of population fluctuation time correlation function of both 4- and 5- coordinated species slows down dramatically and a step-like feature of the relaxation becomes evident, the *lifetime* itself of 5-corrdinated species remains short. These **results suggest a new molecular mechanism of low temperature anomalies and of the liquid-liquid transition in terms of initiation, growth and termination of propagating jump and inter-conversion events.**

As the temperature of water is lowered increasingly below the freezing/melting temperature, it starts to exhibit several startling anomalies which have drawn attention for a long time[1]. Notable among the anomalies is the divergent like growth of the linear response functions (heat capacity, isothermal compressibility, etc)[2,3]. The growth, however, does not continue below a temperature as crystallization intervenes and bulk water can not exist. However, when experiments have been performed on the nanoconfined supercooled water to access the anomalous region as confinement suppresses crystallization, one indeed finds a sharp peak in the response functions, instead of a divergence. In addition to the thermodynamic anomalies, one also finds signatures of fractional diffusion and onset of step-like relaxation in dynamic structure factor[4].

Several explanations of these anomalies have been put forward to explain the above observations. Explanations in terms of re-crossing of the gas-liquid spinodal at lower temperature (spinodal retracing)[5] and in terms of the negative slope of the temperature maximum density line in the ($P,T$) plane (singularity free scenario)[6] are only partly successful in explaining the experimental results. The anomalous increase in the response functions upon cooling at the ambient pressure has recently been attributed to a manifestation of the crossing of the Widom line[7]. While this hypothesis seems at present to be consistent with experiments[8], one can not rule out the existence of a first order phase transition between the high density amorphous and the low density amorphous phases. Several other studies have revealed the relationship between structural order and anomalies of liquid water[9,10], although *macroscopic explanations do not offer satisfactory microscopic picture of the anomalous behavior accompanying the liquid-liquid transition.*

Here we report the results of computer simulations and theoretical analysis which connect the fluctuation between the two forms of water and the response function anomalies. The dynamical origin of the fluctuation is presented. The growth of the response function has been correlated with the growth in the non-linear dynamical response function ($\chi_4(t)$) and which in turn is found to be correlated with the appearance of string-like motion that connects the consecutive inter-conversions between 4- and 5-coordinated water molecules. While several fluctuations show divergent-like behavior near the anomalous region, the fluctuation in the total dipole moment in the system shows the signature of a second order phase transition, in the Ehrenfest sense. We have also noticed that the results presented here bear similarities with the isotropic-nematic phase transition in thermotropic liquid crystals that shows a weakly first order behavior. All these phenomenon can also be explained in terms of an avoided criticality[11] as the growth of long range correlation in 4-coordinated network structure is frustrated due to the presence of 5-coordinated species.

We have identified and categorized the events of transition between a four to five coordinated water molecules and vise versa. The following characteristics features have been observed.

(i) The number fluctuation of 3- and 5-codianted water molecules becomes clearly anti-correlated with the 4-coordinated species in the cross-over region, as shown in **Figure 1(a)**. This result is found to be consistent with the mechanism of the sequential interconversion of the 5-coordinated species to 4-coordinated species in the supercooled water, as will be discussed next.

(ii) In the propagation scenario, a 4-coordinated water molecule gains an extra neighbor from the second solvation shell and becomes 5-coordinated. The 5-coordinated water molecule, however, is short lived and has on the average a lifetime of 1 ps or so. The decay of the 5-coordinated water molecule occurs by the departure of a water molecule from the first solvation shell. **Figure 1(b)** provides a schematic illustration of the event.

(iii) The transition of the central water molecule from the 4- to 5- and again back to 4-corrdianted state is inevitably accompanied by a large amplitude jump in rotation and also in translation[12]. The rotational jump has an average amplitude of about 60 deg. The waiting time distribution of jumps of a tagged water molecule is exponential at high temperatures, but becomes at least bi-exponential as temperature is lowered below the freezing/melting temperature of water. This may lead to fractional diffusion observed in supercooled water.

(iv) These string-like motion ends at a point where the propagating 5-coordination number meets a 3-coordinated water molecule which acts as sink, as illustrated in **Figure 1(b).** However, this string-like motion is quite different from the one observed in the supercooled Lennard-Jones liquid near glass transition[13].

(v) The transformational events become increasingly correlated with each other as the temperature is increasingly lowered in the supercooled region. The bond breaking and forming events take the form of a connected string at low temperaturesThe number of water molecules (n) involved in such a string of correlated hydrogen bond breaking event increases with decreasing temperature. This can be explained as when the temperature is lowered the number of sink (three coordinated water molecules) becomes sharply less (as shown in **Figure 1(c)**). So, the string becomes longer as temperature is lowered in the supercooled region. In **Figure 1 (d),** the histogram of the number of water molecules (*n*) involved in such a string at different temperatures has been presented. It is apparent from the figure that as temperature decreases the population of the number of longer strings increases. Thus, the string-like motion that connects the events of the inter-conversion between four and five coordinated species provides a measure of the dynamical correlation length in the supercooled liquid water.

(a)

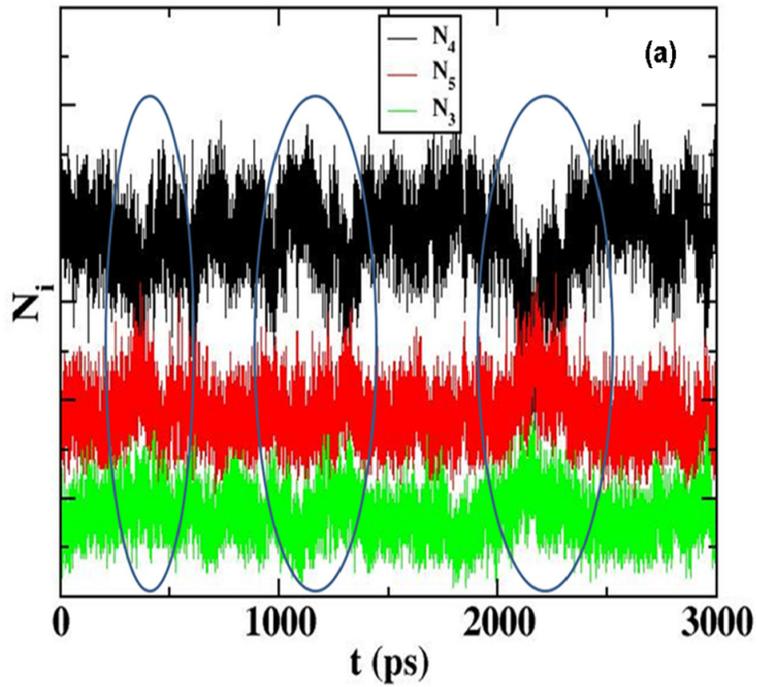

**(b)**

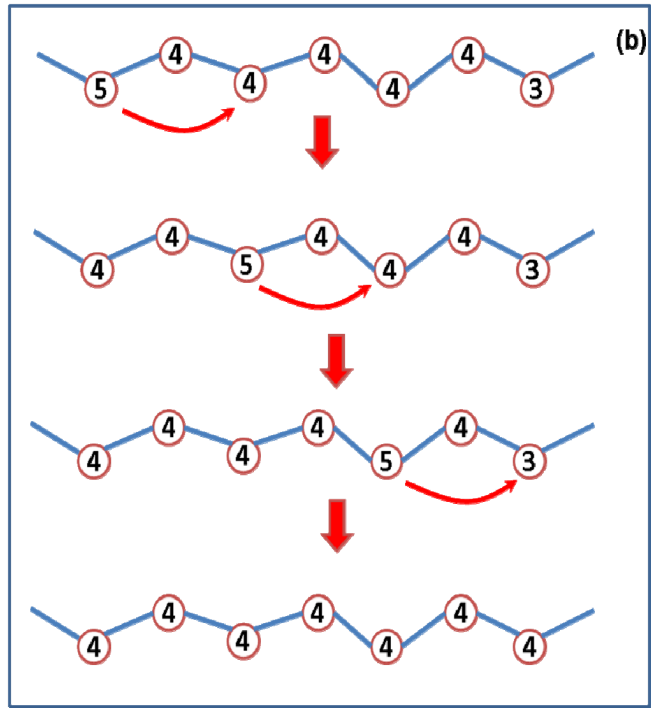

**(c)**

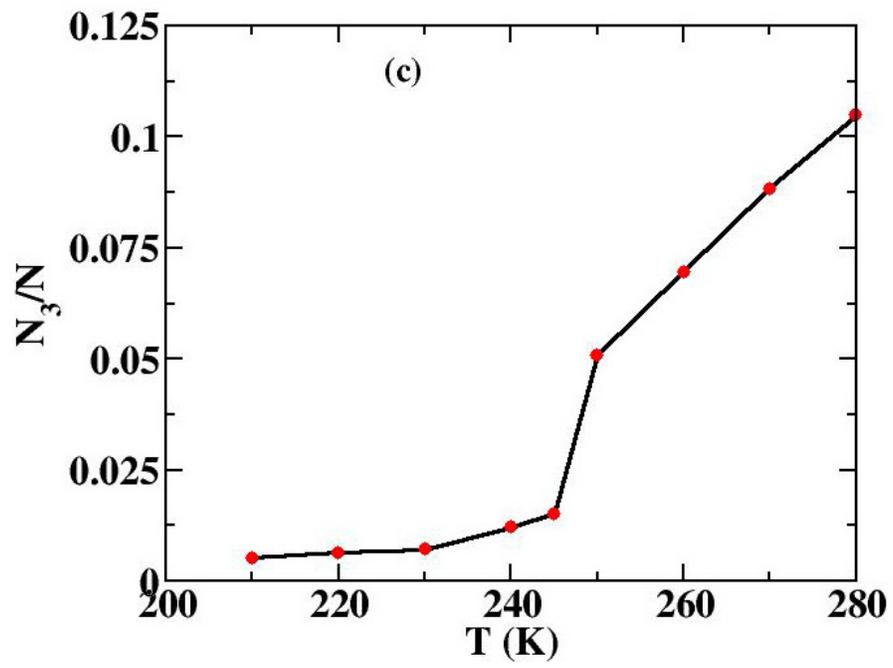

**(d)**

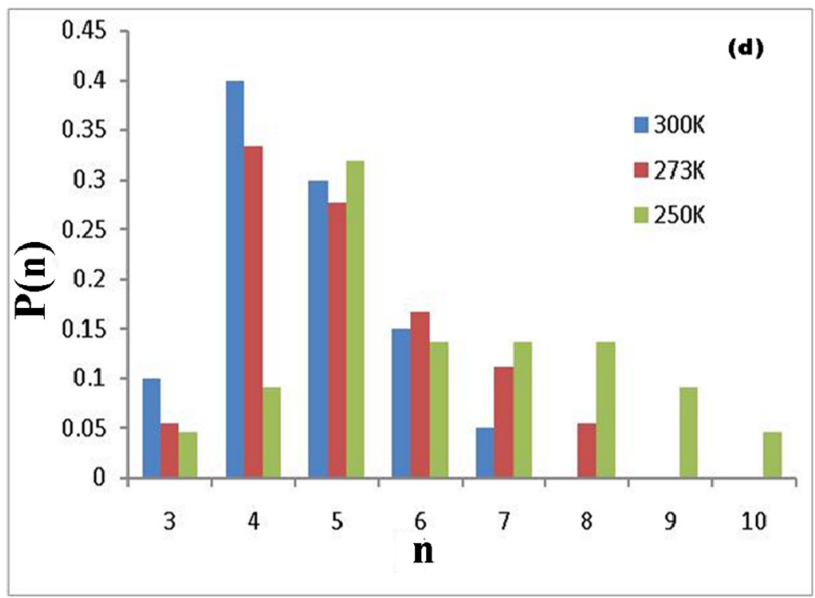

**Figure 1:** (a) Fluctuation in the population of the different species of water at T= 250 K. Note the *in phase* variation between the population of 3- and 5- coordinated species and *out of phase* variation of the population of 4-coordinated species with them. (b) Schematic representation of the mechanism of propagation and death of the string-like motion in supercooled liquid water. The numbers indicated here are the coordination states of the water molecules. The arrow indicates the transfer of water molecules during the H-bond breaking events. For clarity, we have not shown all other water molecules inside the coordination sphere. (c) Fraction of 3-coordinated water molecules in the system in supercooled liquid water. Note the phase transition like behavior at 250 K. (d) Histogram of the distribution of correlation lengths in terms of number of water molecules (n) involved in sequential H-bond breaking events at three different temperatures (300 K, 273K and 250 K).

Some dynamical features of the temperature dependence of the response functions of supercooled water exhibits similarity to supercooled liquids[14] that suggests the emergence of a dynamic correlation length in the low temperature liquid. To quantify this, we have calculated the non-linear response function, $\chi_4(t)$, which measures the dynamical heterogeneity quantitatively. The nonlinear response function, $\chi_4(t)$, is related to the four-point density correlation function $G_4$ by the following relation[15],

$$\chi_4(t) = \frac{\beta V}{N^2} \int dr_1 dr_2 dr_3 dr_4 w(|r_1 - r_2|) w(|r_3 - r_4|) \times G_4(r_1, r_2, r_3, r_4, t) \qquad (1)$$

where four-point density correlation function $G_4$ can be written as[15],

$$\begin{aligned} G_4(r_1, r_2, r_3, r_4, t) &= \langle \rho(r_1, 0) \rho(r_2, t) \rho(r_3, 0) \rho(r_4, t) \rangle \\ &\quad - \langle \rho(r_1, 0) \rho(r_2, t) \rangle \times \langle \rho(r_3, 0) \rho(r_4, t) \rangle \end{aligned} \qquad (2)$$

Thus, $\chi_4(t)$ is dominated by the range of spatial correlation between the localized particles in the fluid. It can be shown that definition of $\chi_4(t)$ as given by Eq. 1 is equivalent to the following expression[15].

$$\chi_4(t) = \frac{\beta V}{N^2} \left[ \langle Q^2(t) \rangle - \langle Q(t) \rangle^2 \right], \qquad (3)$$

where $Q(t)$ is the time dependent order parameter which measures the localization of particles around a central molecule through a overlap function which is unity inside a region *a* and zero otherwise[15]. **Figure 2a** displays the non-linear response function, $\chi_4(t)$, of water molecules (computed with the oxygen atom displacement) for temperatures ranging form *300 K* down to *230 K*. At ambient pressure the linear response function anomalies are located around $T \approx 250K$ for TIP5P water model. For every temperature, the $\chi_4(t)$ is nearly zero for very short and long times with a maximum value at some intermediate time $t^*$. Both $t^*$ and the amplitude of the peak, $\chi_4^M(t^*)$, increase sharply while approaching the anomalous region from the above. This behaviour is similar to what is observed for supercooled liquids closed to glass transition. However, after crossing the anomalous region, the value of $\chi_4^M(t^*)$ decreases with further decrease in temperature. This suggests that both the spatial correlation between slow moving water molecules and the time at which that correlation is maximum, increase as we approach the anomalous region from both the side[10]. **Figure 2b** shows the variation of $\chi_4^M(t^*)$ as a function $t^*$. It exhibits a sharp maximum at $T_{WL} \approx 250K$, indicating that the spatial correlation attains the maximum values at $T= 250\ K$. We also find a direct correlation between the growth in the non-linear response function (peak height of $\chi_4(t)$) and anomalous behavior of the thermodynamic response function ($C_P$, $\kappa_T$, etc.), as shown in **Figure 2(c)**. Thus one can explain the anomalies in supercooled water as a consequence of the growth behavior of $\chi_4(t)$ across $T = 250K$.

**(a)**

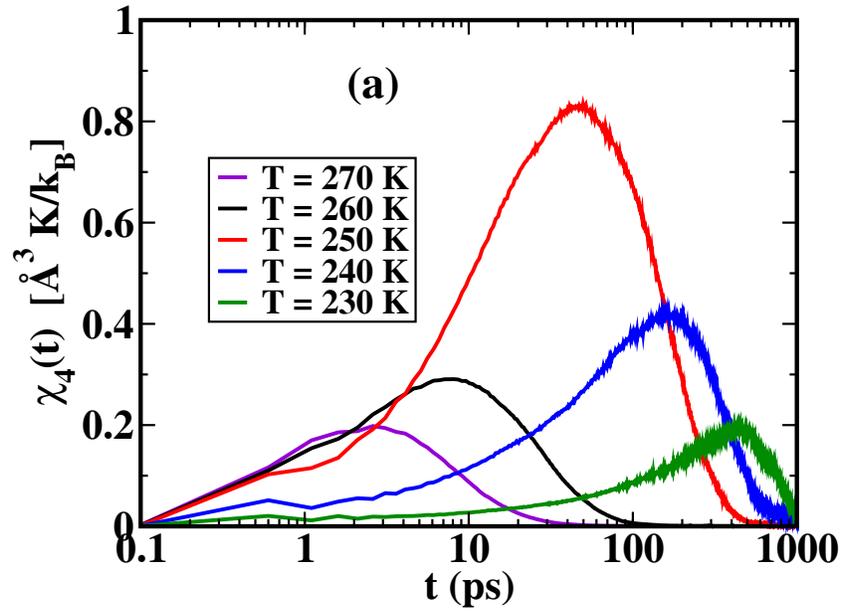

**(b)**

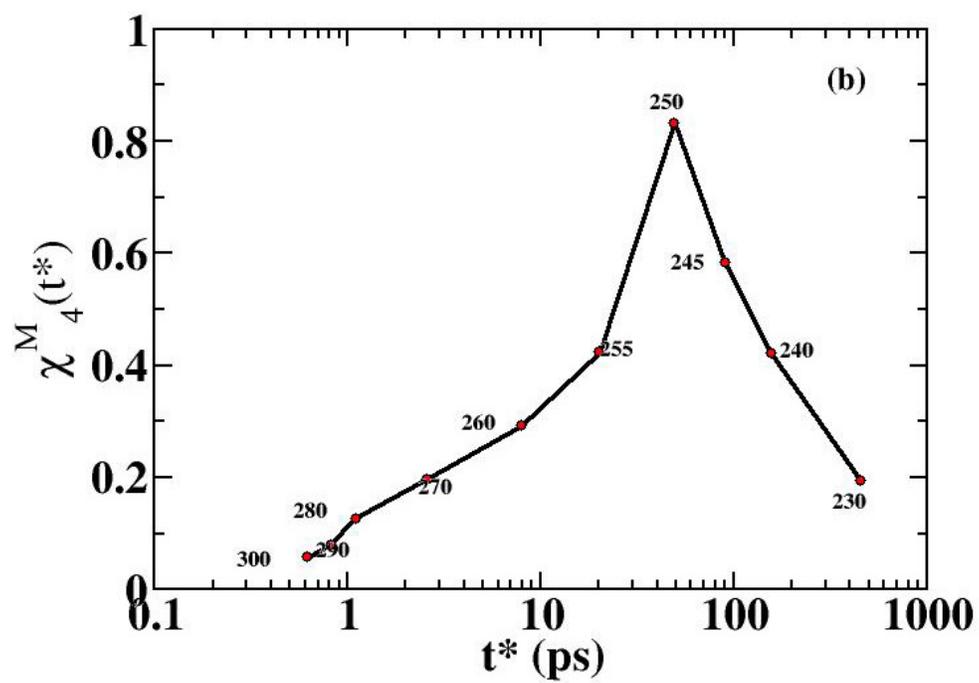

**(c)**

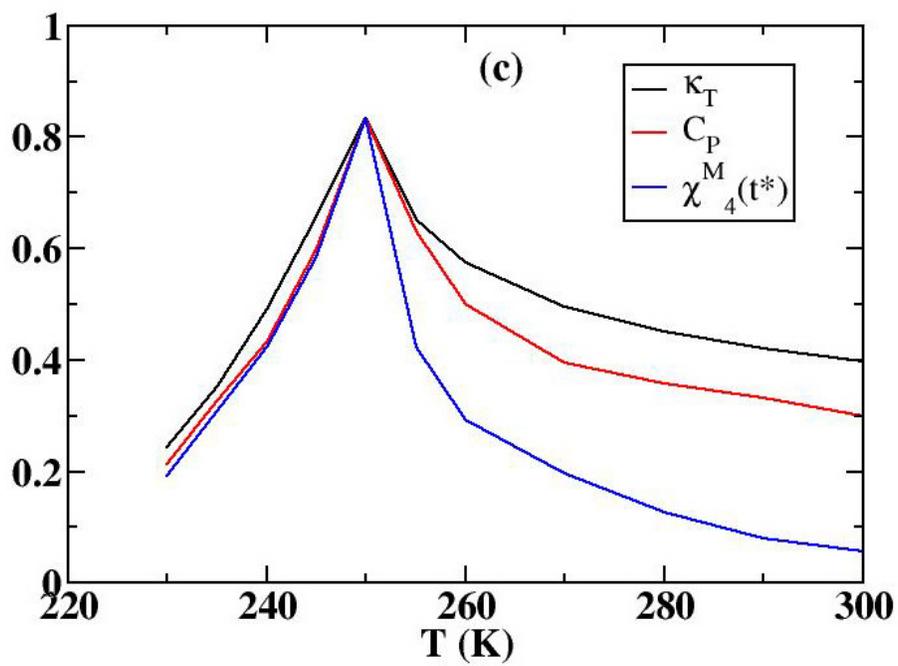

**Figure 2:** (a) Time dependence of non-linear dynamic response function, $\chi_4(t)$, at several temperatures starting from 300K down to 230K in supercooled liquid water. Note the huge peak at *T= 250K*. (b) Variation of the maximum of $\chi_4(t)$, $\chi_4^M(t^*)$, as a function of the time at which $\chi_4(t)$ attain maximum, $t^*$. Note the maximum at T= 250K. (c) Variation of $\kappa_T$, $C_P$ and $\chi_4^M(t^*)$ with temperature in the supercooled water. Note that all three quantities attain maximum value at T = 250 K.

Variation of orientational correlation across the transition region could be a good marker of the nature of the transition. Therefore, we calculated the fluctuation of the mean square of the total dipole moment $M(t) = \sum_i \mu_i(t)$ of the system with temperature. In **Figure 3** we show $\langle(\delta M)^2\rangle$ as a function of temperature. A sharp decrease in the value of $\langle(\delta M)^2\rangle$ is observed as the system crosses T= 250 K from above. This sharp decrease is also reflected in the value of static dielectric constant of water which shows a fall from ~60 to ~7 as it crosses T = 250 K. Thus, the onset of decrease in the $\langle(\delta M)^2\rangle$ or static dielectric constant of water specifies the region of thermodynamic anomalies. This observation explains several reports on the dramatic change in the hydrophobic hydration upon crossing 250 K. This variation is clearly connected with a sharp change in the curvature (or, force constant) of the free energy surface along total dipole moment fluctuation. The low temperature LDL phase is having a larger curvature, characteristics of a strong liquid. Since $\langle(\delta M)^2\rangle$ is given by the second derivative of the free energy, its sharp variation seems to suggest a second order phase transition in the Ehrenfest sense or even a weak fist order phase transition.

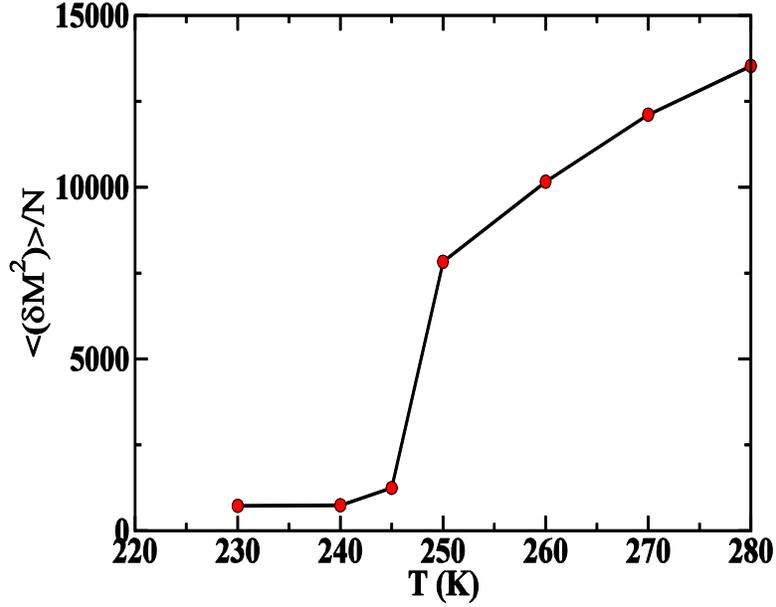

**Figure 3:** Variation of $\langle(\delta M)^2\rangle/N$ of the system across the anomalous region. M is the total moment of the system and N is the number of water molecules present in the system. Note the sharp decrease in $\langle(\delta M)^2\rangle/N$ upon crossing T=250 K from the above.

Next, we investigate the correlation of the number fluctuation of the *4-* and *5*-coordinated species in time. The correlation function $C_i(t)$ (*i = 4, 5*) is defined as,

$$C_i(t) = \frac{\langle \delta N_i(0) \delta N_i(t) \rangle}{\langle \delta N_i(0) \delta N_i(0) \rangle}, \qquad (4)$$

**Figures 4(*a*)** displays the decay of the number fluctuation time correlation function in log-log plots for *5-* coordinated species. The decay of the self fluctuation time correlation function of 5-coordinated species (as shown in **Figure 4(*a*)** becomes slowest at *T = 250 K* (anomalous region). A step-like feature also develops (similar to the relaxation in supercooled liquids and thermotropic liquid crystals[14,16]) in the relaxation pattern as the anomalous region is approached from above and continues to be present after crossing. The decay pattern of the 4-coordinated species is found to be similar to the 5-coordinated species as their fluctuations are anti-correlated

in time[10]. The slowest decay near the anomalous region ($T = 250$ K) suggests the appearance of strong anti-correlation with slow decay time. The appearance of step-like feature in the relaxation pattern of self and cross fluctuation time correlation function of the species are the indication of the crossing of anomalous region for the supercooled water.

We next discuss the lifetime dynamics of the different species. The lifetime correlation function $S_{Li}(t)$ ($i = 4,5$) of *4-* and *5-*coordinated water molecules is defined as,

$$S_{Li}(t) = \frac{\langle n_i(0) n_i(t) \rangle}{\langle n_i(0) \rangle}, \qquad (5)$$

where $n_i(t)$ is considered to be unity if a water molecule is continuously found in a state of coordination number $i$ for the time interval *0* to *t*, and otherwise zero. **Figures 4b** displays the decay of the correlation functions and corresponding temperature dependent lifetimes of *4-* and *5-*coordinated species. For the *4-*coordinated species, the decay of the lifetime correlation function $S_{L4}(t)$ (upper left panel) markedly slows down upon cooling across the anomalous region. This is similar to features observed in the dynamics (such as in dynamic structure factor, $F_S(k,t)$[14]) of supercooled water. On the other hand, the decay of $S_{L5}(t)$ (upper right panel) remains fast over the entire temperature range and shows no dramatic change on crossing the anomalous region. We have also calculated the average lifetime by fitting of the correlation functions. The result is displayed in the lower panel of **figure 4b.** This indicates that the barrier height for the decay of the *4-*coordinated species becomes higher after crossing the anomalous region, but the same for *5-*coordinated species continues to remain low.

**(a)**

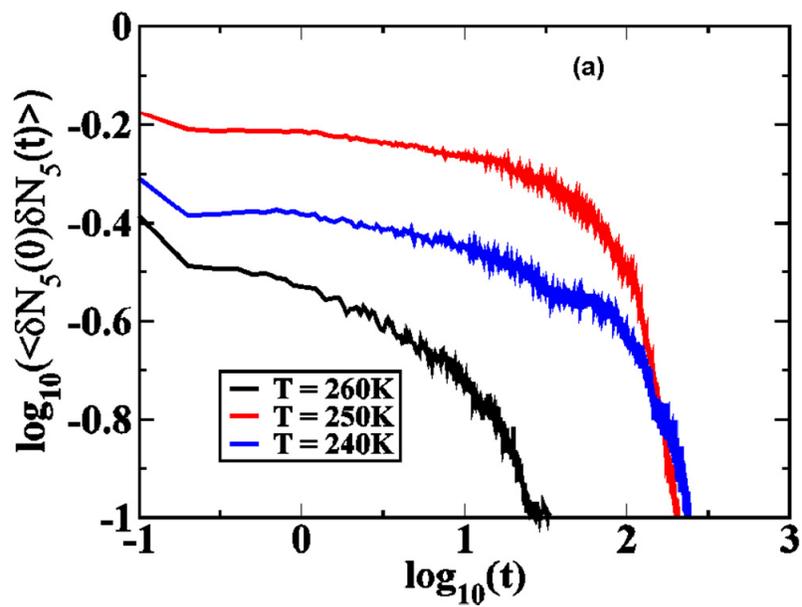

**(b)**

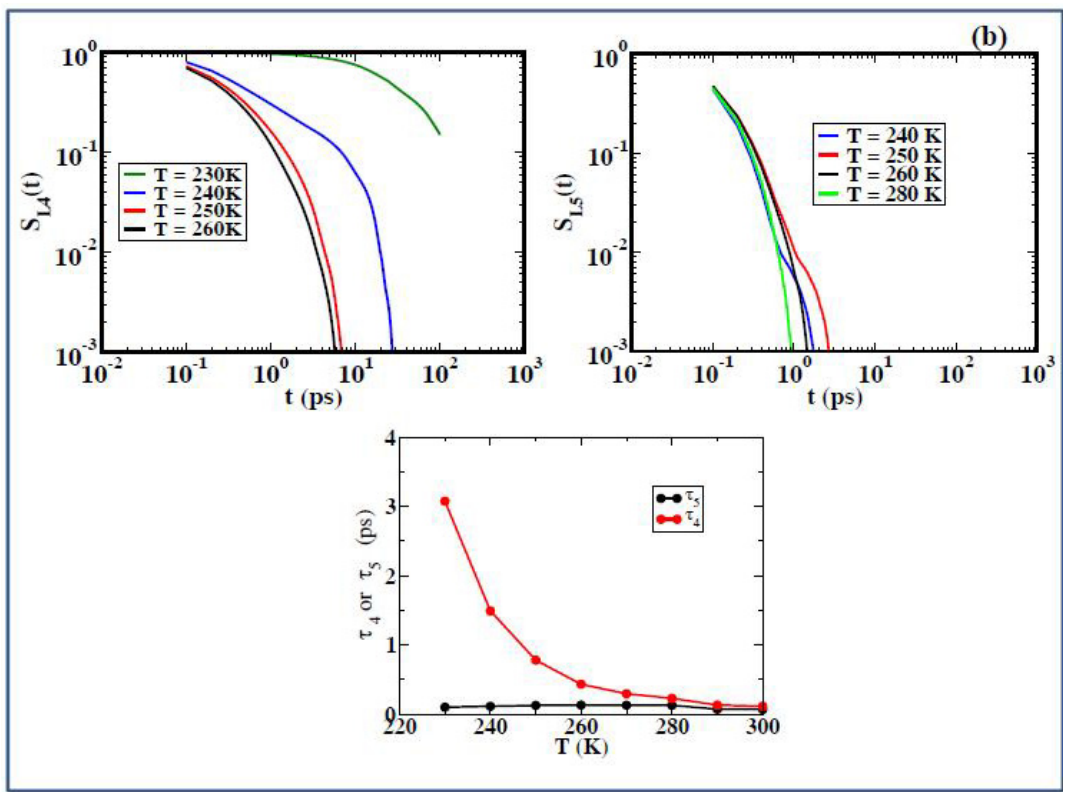

**Figure 4** (a) Decay of fluctuation time correlation function, $C_5(t)$ in log-log plot. Note the slowest decay of al the correlation functions at $T = 250\ K$ and also the emergence of step-like feature near the anomalous region. (b) Contrasting lifetime decay behaviour of 4- (upper left panel) and 5-corrdinated (upper right panel) species and the temperature variation of the corresponding lifetimes (lower panel).

The following molecular mechanism emerges from the present analysis. As the temperature of water is lowered below the freezing/melting temperature, the fraction of both 3- and 5 – coordinated water molecules continue to decrease while that of 4-coordinated water molecules increases. While inter-conversions between these molecules continue due to thermal fluctuations, they become more correlated, as the rigidity of the hydrogen bond network increases. As a 5-coordinated water molecule undergoes a change to a 4-coordinated one, another water molecule in the second nearest neighbor becomes 5-coordinated. This synergy between inter-conversions gives rise to a string-like connectivity which can be terminated only when the string meets a 3-coordinated water molecule which becomes 4-coordinated. A schematic representation of the mechanism is presented in **Figure 1a.** As the fraction of 3-coordinated water molecules decreases with lowering temperature and the rigidity of the hydrogen bond network increases, the average length of the string increases. Since the creation of a 5-coordinated water molecule (by thermal fluctuation) accounts to a density fluctuation, the same fluctuation can propagate and can create other strings. This gives rise to the observed intermittency in the time dependency of 4- and 5-coordinated water molecules whose number fluctuations are anti-correlated. As expected, we find that the decrease in the number of 5-coordinated water molecule occurs in phase with the number of 3-coordinated water molecules and that is found to be accompanied by an increase in the 4-coordinated species in the system (as shown in **Figure 1c**), thus supporting the argument put forward here.

The anti-correlated fluctuations between the number of 4-coordiated water molecules on one side and 3- and 5-coordinated water molecules on the other are responsible for the sharp divergent-like rise in specific heat and isothermal compressibility, and also a sharp change in the mean-square dipole moment fluctuations. Across 247 K, the dielectric constant of water changes sharply, as fluctuations decrease. The characteristics exhibited by the system are the one expected for a weakly first order phase transition. We find weak dependence on the system size.

In order to obtain the dependence of the lifetime of the string on the length of the string, we model the motion of the 5-coordinated water molecules (excitation) as a 1-dimensional random walk between two absorbing boundaries, formed by 3-coordinated water molecules (see figure 1(b)). The survival probability of the 5-coordinated species can be obtained using the method of images[17]. The final expression for the survival probability is given by,

$$S_5(t) = \frac{4}{\pi}\exp(-k_0 t) \times \sum_{n=0}^{\infty} \frac{1}{2n+1}\exp\left[-\frac{(2n+1)^2 \pi^2}{l^2}Dt\right] \times \int_{-\infty}^{\infty} dx' P_0(x')\sin\frac{(2n+1)\pi x'}{l} \quad (6)$$

where $k_0$ is the rate constant of the decay of 5-coordinated water molecules by the processes (other than death by interaction with 3-coordinated water molecules), $l$ is the length of the string., and $P_0(x)$ is the initial position of the tagged 5-coordinated water molecule.

As the number of 3-coordinated water molecules ($<N_3>$) decreases (sharply) with decreasing temperature (particularly near 250 K, as shown in figure 1(c)), the length of the string should increase ($l \sim <N_3>^{-1/3}$) and the survival time of the 5-coordinated species should increase as $\tau_s^5 \sim <N_3>^{-2/3}$. At around 250K, different strings survive for longer time. The strings then interact with each other leading to a phase transition like behavior mentioned above. If we increase temperature from the low temperature low density liquid (LDL) phase, then the transition to the high density (HDL) liquid phase occurs through the formation of these strings which leads to the breakdown of the hydrogen bond network, in a fashion quite similar to the melting transition of a solid. In fact, a recently proposed scenario invokes proliferation of strings to drive a glass transition which bears similarity to the present study[18].

Below 270K, the relaxation in water is found predominantly to occur through orientational jump events accompanying conversion of a 4-coordinated water molecule to a 5-coordinated one and vise versa. In such events one of the water molecules (the incoming 5$^{th}$ neighbor during 4- to 5-coordination transition) undergoes translational displacement. The jump event as shown previously can lead to significant relaxation of the local stress[19]. However, only one water molecule undergoes significant translational displacement per event. Such discrete events are known to provide deviation from the Stokes-Einstein relationship between self diffusivity and liquid viscosity with an observed value of diffusion coefficient *smaller* than the prediction of Stokes-Einstein relationship.

To summarize, we find a growth of string-like dynamically connected motion of coordination number *five* in the supercooled liquid water upon decreasing temperature which originates from the sequential inter-conversion between four and five coordinated water molecules. The formation of the string is found to be facilitated by the small lifetime of the 5-coordinated species and small fraction of 3-coordinated species in supercooled water. The presence of such strings in a few locations induces a divergent-like growth of the non-linear dynamical response function precisely at a temperature where all the thermodynamic response functions show anomalies. Finally, we have observed similar propagation of jump events also in the model SPC/E water at low temperature, and therefore expect them to be the dominant dynamical process in supercooled water.

**Methods**

A series of eleven simulations at different temperature starting from *300 K* to *230 K* has been carried out. The water model used is *TIP5P*[20] which has been examined to reproduce the low temperature properties of water excellently. A system of *512* water molecules is first equilibrated for several nanoseconds (around *1* ns for *300 K* and around *15* ns for *230 K*) in isobaric isothermal (NPT) ensemble to reproduce the correct density and then trajectory is saved for another several nanoseconds for the analysis in canonical (NVT) ensemble. For thermodynamics response function calculation, after equilibration the trajectory is saved for several nanosecond in NPT ensemble. The long range electrostatic interactions are calculated using Ewald summation.

**Acknowledgements**

We thank Prof. G. Ananthakisna and Prof. P. G. Wolynes for helpful discussion. This work has been supported in parts by the grants from DST and JC Bose Fellowship, India. BJ thanks CSIR, India for providing research Fellowship.